\documentclass[submission,copyright,creativecommons]{eptcs}

\providecommand{\event}{Post Proceedings of ThEdu'17}

\usepackage[english]{babel}
\usepackage[utf8x]{inputenc}
\usepackage[T1]{fontenc}

\usepackage{graphicx}
\usepackage{amssymb} 

\usepackage{listings}

\lstdefinelanguage{JSON}
{morekeywords={module,signature,axioms,view,sort,protecting,op,eq,var},
  sensitive=false,
  showspaces=false,
  showstringspaces=false
  morecomment=[l]{--},
  morestring=[b]",
}

\newcommand{\wglversion}{{1.4}}
\newcommand{\WGL}{\textit{WGL\/}}
\newcommand{\cpp}{\textsc{C$++$}}
\newcommand{\itog}{{\sc i2g}}
\newcommand{\itogatp}{{\sc i2gatp}}

\newcommand{\TGTP}{\textit{TGTP\/}}

\title{Exchange of Geometric Information Between Applications}

\author{Pedro Quaresma
  \institute{
    CISUC / Department of Mathematics \\
    University of Coimbra, Portugal} \\
    \email{pedro@mat.uc.pt}
    \and
    Vanda Santos
    \institute{CISUC\\
      University of Coimbra, Portugal} \\
    \email{vsantos7@gmail.com} \\
    \and
    Nuno Baeta
    \institute{
      CISUC\\
      University of Coimbra, Portugal} \\
      \email{nmsbaeta@gmail.com} \\
}

\def\authorrunning{P. Quaresma, V. Santos \& N. Baeta}

\def\titlerunning{Exchange of Geometric Information Between
  Applications}

\begin{document}

\maketitle

\begin{abstract}
  The Web Geometry Laboratory ({\WGL}) is a collaborative and adaptive
  e-learning Web platform integrating a well known dynamic geometry
  system. Thousands of Geometric problems for Geometric Theorem
  Provers ({\TGTP}) is a Web-based repository of geometric problems to
  support the testing and evaluation of geometric automated theorem
  proving systems.

  The users of these systems should be able to profit from each
  other. The {\TGTP} corpus must be made available to the {\WGL} user,
  allowing, in this way, the exploration of {\TGTP} problems and their
  proofs. On the other direction {\TGTP} could gain by the
  possibility of a wider users base submitting new  problems.

  Such information exchange between clients (e.g. {\WGL}) and servers
  (e.g. {\TGTP}) raises many issues: geometric search---someone,
  working in a geometric problem, must be able to ask for more
  information regarding that construction; levels of geometric
  knowledge and interest---the problems in the servers must be
  classified in such a way that, in response to a client query, only
  the problems in the user's level and/or interest are returned;
  different aims of each tool---e.g. {\WGL} is about secondary school
  geometry, {\TGTP} is about formal proofs in semi-analytic and
  algebraic proof methods, not a perfect match indeed; localisation
  issues, e.g. a Portuguese user obliged to make the query and process
  the answer in English; technical issues---many technical issues need
  to be addressed to make this exchange of geometric information
  possible and useful.

  Instead of a giant (difficult to maintain) tool, trying to cover
  all, the interconnection of specialised tools seems much more
  promising. The challenges to make that connection work are many and
  difficult, but, it is the authors impression, not insurmountable.
 \end{abstract}

\section{Introduction}
\label{sec:introduction}

In the area of geometry there are now a large number of computational
tools that can be used to perform many different tasks, dynamic
geometry systems (DGS), computer algebra systems (CAS), geometry
automatic theorem provers (GATP) among
others~\cite{Quaresma2017}. These tools are being used in many
different settings, in education, in research, etc. Apart from the
set of examples that can be found in many of the systems currently
available, there are several repositories of geometric knowledge
(e.g. \emph{Intergeo},\footnote{\href{http://i2geo.net/?language=pt}{The
    Interoperable Interactive Geometry for Europe.}}
{\TGTP},\footnote{\href{http://hilbert.mat.uc.pt/TGTP/index.php}{Thousand
    of Geometric problems for geometric Theorem Provers.}}
\emph{GeoGebra
  Materials}\footnote{\href{https://www.geogebra.org/materials/}{GeoGebra
    Materials.}}).

When considering the current status, two questions arise:

\begin{itemize}
\item How systems that use geometric information and geometric
  information repositories can be put to work together?
\item Should they be put to work together?
\end{itemize}

The answer to the second question is, in the authors opinion:
\emph{Yes}. For users of a system that manipulates information, the
access to repositories of geometric information (problems,
conjectures, proofs, etc.) will be very interesting. It will
constitute a wide source of examples to work with, enlarging the
(eventual) set of its own examples. For a collaborative build
repository of geometric information enlarging its users/contributors
base will be useful, thus enabling to open the repository to more, and
of different types, contributions.

The answer to the first question lies on the answers to problems
raised by the interconnection of those type of systems: search
mechanisms (see Section~\ref{sec:geometricsearch}); adaptive filtering
of the geometric information (see
Section~\ref{sec:adaptivefiltering}); collaborative editing (see
Section~\ref{sec:collaborativeediting}); technical issues raised by
the geometric information interchange (see
Section~\ref{sec:TechnicalIssues}).

\textit{Overview of the paper.} The paper is organised as follows:
first, in Section~\ref{sec:wglmeetstgtp}, we introduce our test-case,
the interconnection between the systems {\WGL} and {\TGTP}. In
Section~\ref{sec:IssuesintheWGLTGTPinterconnection} we analyse the
different problems and solutions in making the interconnection between
clients and servers of geometric information. In
Section~\ref{sec:conclusions} final conclusions are drawn.

\section{Clients and Servers of Geometric Information}
\label{sec:wglmeetstgtp}

In this section two specific systems will be introduced. These systems
are representative of the clients and servers systems in need of being
connected. This will give an actual setting where the interconnection
can and should be implemented.

The Web Geometry Laboratory ({\WGL}) is a collaborative and adaptive
e-learning Web platform integrating a well known
DGS~\cite{Quaresma2017a,Santos2018}. Thousands of Geometric problems
for geometric Theorem Provers ({\TGTP}) is a Web-based repository of
geometric problems to support the testing and evaluation of geometric
automated theorem proving systems~\cite{Quaresma2011}.

The interconnection of the two systems, {\WGL} and {\TGTP}, will
reinforce the usefulness of each other.

The {\WGL}$\rightarrow${\TGTP} interconnection will provide to the
{\WGL} users a large set of geometric constructions, giving them the
possibility of browsing constructions and exploring conjectures and
proofs. Teachers and students will gain the possibility of visualising
geometric problems and link them with formal
proofs~\cite{Quaresma2016}.

The study of proofs in education is important as can be seen by the
many contributions in proceedings of the ICMI Study 19 Conference:
\emph{Proof and Proving in Mathematics Education}~\cite{Hanna2012}.  A
number of DGS already incorporates GATPs: \emph{GCLC} incorporates
four GATPs in it~\cite{Janicic2006c}; new versions of
\emph{GeoGebra}~\cite{Hohenwarter2002} already include a connection to
GATPs allowing to give a formal answer to a given validation
question~\cite{Botana2015a}; \emph{Java Geometry
  Expert}~\cite{Ye2010a,Ye2010b,Ye2011a} incorporates several
GATPs. Cinderella uses a technique called ``randomised theorem
checking'', generating a large number of random examples, checking if
the conjecture holds, establishing the truthfulness if the answer is
yes for all examples generated~\cite{Richter-Gebert2000}.

None of above mentioned systems possess a repository of problems (some
of them have unstructured lists of examples) open to users
queries. The interconnection of clients and servers of geometric
information will provide users with that open access to geometric
information.

On the other hand, the {\WGL}$\leftarrow${\TGTP} connection will allow
to enlarge the users base of {\TGTP}: teachers, and eventually
students, could submit new conjectures. This will contribute to
{\TGTP}'s overall goal of providing a comprehensive repository of
geometric problems, but will also enlarge its usefulness in education,
given the fact that it is expected that secondary school teachers, and
students, will contribute with problems close to the geometric
subjects they are studying.

\section{Issues in the Clients/Servers Interconnection}
\label{sec:IssuesintheWGLTGTPinterconnection}

As in the case of {\WGL}$\leftrightarrows${\TGTP} interconnection, the
generic interconnection of clients and servers of geometric
information raises many issues:

\begin{description}
\item[Searching:] having repositories of geometric knowledge, users
  should be able to query the information in the most fruitful and
  easy way.
\item[Adaptive Filtering:] different users will have different goals
  when addressing a repository of geometric information: the user's
  level of geometric knowledge; the user's learning profile; the
  intended use of the geometric knowledge (e.g. educational setting,
  automatic deduction research).
\item[Collaborative Editing:] to allow repositories of geometric
  knowledge to grow, a collaborating editing interface, appealing and
  easy to use, should be implemented. The information introduced
  should contain keywords and other semantic information, allowing an
  easy querying.
\item[Localisation:] the {\WGL} is an internationalised system, i.e. it
  has English as the base language and the Portuguese and Serbian
  translations (version \wglversion). The {\TGTP} is an English
  language only system. Setting aside any form of automatic
  translation, in this article that problem will not be approached,
  this is again an adaptive filtering issue, knowing the users
  preference the problems in the appropriated language(s) will be
  selected.
\item[Technical issues:] to make a connection between two different
  types of systems some technical issues must be addressed. The main
  ones are: specification of application interface protocols; common
  format for geometric information.
\end{description}

In the following these issues will be addressed.

\subsection{Searching for Geometric Information}
\label{sec:searchinggeometricinformation}


Having a repository of geometry knowledge a user should be able to
browse the information in an easy and useful way. To a collaboratively
built repository it is also important to avoid repetition and ensure
consistency (e.g. non-contradictory facts). To answer both issues good
search mechanisms are needed.

In \href{https://www.geogebra.org/search}{\emph{GeoGebra
    materials}}\footnote{In 2018-01-11 there were 1041831 materials.}
the search ``Ceva'' returned more than one hundred answers, most of
them are about the same subject, ``Ceva's Theorem''. So a bit of
excess of collaborative editing (many repetitions) and a simple
text search mechanism. In the
\href{http://i2geo.net/}{\emph{Intergeo} site}\footnote{In 2018-01-11
  there were 3957 resources.}, there is a ``simple text search'' and a
``complex text search''. For the same text search (``Ceva'') we get
three results, two of them about ``Ceva's Theorem'' (a construction
and an applet). The complex text search opens the possibility of
adaptive filtering the information (e.g. choose the type activity:
exercises; game; etc.). In
\href{http://hilbert.mat.uc.pt/TGTP/index.php}{\TGTP\ }\footnote{In
  2018-01-11 there were 236 problems.} using the same simple text
search returns one single result, ``Ceva's Theorem''. Also available
is an extended text search and additionally, a novelty in the
geometric information repositories, a geometric search mechanism,
whose details will be explained below.

The above cited examples have different goals: generic geometric
objects, for \emph{GeoGebra materials} and \emph{Intergeo};
conjectures to be proved by GATPs in {\TGTP}. All systems share a
common simple text search mechanism, \emph{Intergeo} adds to that an
adaptive filtering mechanism, and {\TGTP} a geometric search
mechanism.

A user of such systems will be looking for constructions, problems,
conjectures, geometric information in a generic form. For example,
\emph{The circumcircle of a triangle}, \emph{Ceva's Theorem};
\emph{triangle concurrency points}, among many other possible queries.

\paragraph{Text Search}
\label{sec:textsearch}

The text search mechanisms are applied to one or more of the
attributes that characterise the geometric objects in the
repositories. This search mechanisms can be simple or complex, for
example in {\TGTP} the simple text query is done using \emph{MySQL}
regular expressions~\cite{mysql2011}, over the \texttt{name} field in
the database. A more powerful, text search mechanism is also
available, using the \emph{full-text search} of
\emph{MySQL}~\cite{mysql2011}, over the fields \texttt{name},
\texttt{description}, \texttt{shortDescription} and \texttt{keywords}.

To facilitate the search, the information contained in the
repositories should be enriched, manually or in some automatic way,
with the use of geometric ontologies or with the addition of
meta-information (e.g. keywords) that can characterise the geometric
information~\cite{Chen2012}.


{\TGTP} does not have any automatic process for adding
meta-information to the different entries in its repository. When
inserting a new geometric conjecture the users can insert keywords
characterising the new entry. 


\paragraph{Geometric Search}
\label{sec:geometricsearch}

In geometry, apart from textual approaches common to other areas of
mathematics, there is also the need for a geometric search approach,
i.e. semantic searching in a corpus of geometric constructions.

We should be able to retrieve the geometric information contained in
the many repositories of geometric knowledge. That is, we should be
able to query for a given geometric construction having as result a
set of similar geometric constructions, i.e.  construction with, at
least, the same geometric properties of the query construction.

Based on some preliminary work in geometric
search~\cite{Haralambous2014}, {\TGTP}, apart from the two text
queries mechanisms, has also a geometric search mechanism. The queries
are constructed using a DGS and the geometric construction is
semantically compared with the geometric constructions in the
repository. The result of a query will be a list of geometric
constructions, with the part matching the query highlighted.

To search, semantically, a given geometric constructions, it must be
possible to perform inferential closure, starting with the geometric
construction and obtaining a fix-point, the semantic description of
that geometric construction, with all the properties that can be
extracted (inferred) from it. The search mechanism will proceed by
looking for other geometric constructions with a superset of these
properties. The geometric construction, object of the query, is equal
or a subfigure of those figures.\footnote{Yannis Haralambous and
  Pedro Quaresma, Geometric Figure Mining via Conceptual Graphs in
  preparation.}

The first version of the geometric semantic search mechanism is
already implemented in {\TGTP}, using the \emph{GeoGebra} JavaScript
applet for building the geometric query. The proposed geometric
(semantic) search mechanism implemented in {\TGTP} works as follows:

\begin{enumerate}
\item the user builds the geometric query using \emph{GeoGebra};
\item that geometric construction is transformed from the DGS format
  to a \emph{predicate format};
\item the \emph{predicate format} is converted into a \emph{conceptual
    graph};
\item the inferential closure is performed;
\item the result is converted in a \emph{global trail distribution} (GTD),
  a compact easy to manipulate representation of the conceptual graph;
\item a database query is performed (almost all geometric
  constructions have a corresponding GTD), finding a list of
  all admissible candidates, i.e. constructions whose GTDs are a
  super-set of the query GTD;
\item using a subgraph isomorphism algorithm, identify the query
  construction as a subconstruction of the list of admissible
  candidates.
\end{enumerate}

The last step is still to be implemented in {\TGTP}. For now the
process stops after getting the list of all admissible candidates and
such list constitutes the output of the query.

In the current implementation the quality of the geometric search
approach can be, somehow, misleading. If the user draws, three points,
the lines connecting them and a circle, then the 96 (out of 193)
constructions returned, are about the \emph{incircle of a triangle},
\emph{circumcircle}\footnote{{\TGTP}'s geometric conjectures
  \texttt{GEO0281} and \texttt{GEO0328} respectively.} and many other
constructions involving circles, which is a very good fit (for the
list of admissible candidates). Not all constructions on the previous
geometric search are about triangles, but if we make a geometric
search about three points and the lines connecting them, thus forming
a triangle, the result is not the expected one, we get 193 (out of
193) answers, i.e. no useful selection was made. The reason for this
is that the constructions in the database, and the geometric search
also, are about points and lines, not having the triangle as object,
so the match is made with all the constructions that have, at least,
three points, and three lines. Not surprisingly all the constructions
are more complex than that.

\vspace*{0.6em}

Summarising this section, we can conclude that some sort of
complex text query mechanism (maybe a combination of {\TGTP} and
\emph{Intergeo} text search) is needed, with filtering (see below). A
geometric query mechanism will be also a very important addition.

\subsection{Adaptive Filtering}
\label{sec:adaptivefiltering}


The problems in a repository of geometric information must be
classified in such a way that, in response to a user's query, only the
problems in the user's level and/or learning style and/or interests
and/or language(s) are returned.

\paragraph{Type of Users}
\label{sec:typeusers}

Different type of users will search for different type of information,
e.g. the respective audience is different for {\WGL} and {\TGTP}: the
former is constituted by secondary school teachers and students, the
latter by computer science researchers, experts in geometric automated
theorem proving. The same query made by {\WGL} and by {\TGTP} users
should have different outcomes, adjusted to the user's expectations.
Even if we consider only the {\WGL} users, we should expect many
different search goals, connected to the many different levels of
geometric knowledge~\cite{Crowley1987,Usiskin1982} and also many
different learning
styles~\cite{Brusilovsky1998,Lee2001,Papanikolaou2002}.

\paragraph{Localisation Issues}
\label{sec:localisationissued}

A user of a client service, using a given language (e.g. Portuguese)
may not want to receive the information from a server in a different
language (e.g. English). This may be the case with {\WGL} and {\TGTP},
a secondary school student using {\WGL} in Portuguese may have
difficulties in dealing with an answer, in English, from the {\TGTP}
server. In our view, this is a complex problem and it will not be
addressed in this article.

\paragraph{Taxonomies of Geometric Problems}
\label{sec:taxonomiesgeometricproblems}

Taxonomies of geometric problems must be researched and implemented in
the repositories of geometric knowledge in such a way that the users'
queries can be filtered by a given criteria, adapting the result to
the user.~\footnote{Pedro Quaresma, Vanda Santos, Pierluigi Graziani
  and Nuno Baeta, \emph{Taxonomies of Geometric Problems}, to be
  submitted to a special issue of Journal of Symbolic Computation, on
  Dynamic Geometry and Automated Reasoning, full paper version of a
  extended abstract accepted at ThEdu'17.}

\paragraph{Current Status and Next Steps}
\label{sec:currentstatusandnextsteps}

\begin{figure}[hbtp]
  \centering
      \includegraphics[width=0.8\textwidth]{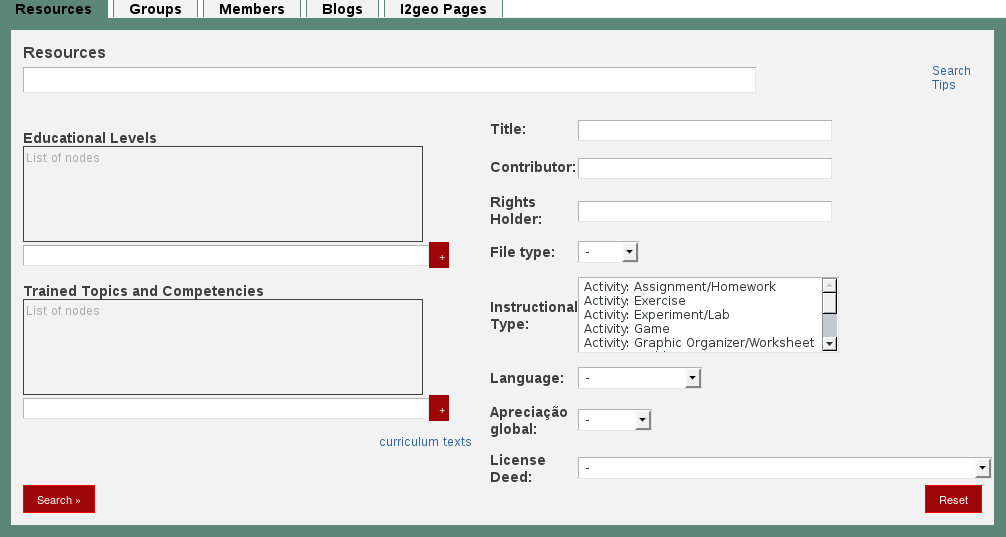}
  \caption{I2G Advanced Search Mechanism}
  \label{fig:I2GAdvancedSearchMechanism}
\end{figure}

A first approach to adaptive filtering can be seen in the
\emph{Intergeo} project's ``advanced search'' mechanism. It has some
adaptive filtering options, ``Educational Levels'' and ``Trained
Topics and Competencies'', can be considered among others filters (see
Figure~\ref{fig:I2GAdvancedSearchMechanism}). This allows a better fit
to the different users needs.

The next step should be to implement adaptive features in such a way
that the system would be capable of retaining information about the
user, applying filters in an automatic
way~\cite{Chrysafiadi2013,Shute2001,Triantafillou2003}. As far as we
know none of the current systems has this feature.

\subsection{Collaborative Editing}
\label{sec:collaborativeediting}

When building a repository with knowledge form a given area is now
common practice to do it in a collaborative way, that is, open the
manipulation of information contained in the repository to a large
base of collaborators, enabling them to insert, update and delete
``objects'' of information.

All the cited repositories (see Section~\ref{sec:introduction}) work
that way. The problem is how to avoid redundancy and incongruity. As
the example in section~\ref{sec:searchinggeometricinformation} showed,
we can fill a repository with many redundant information, reducing its
usefulness. An even worst situation can occur when contradictory pieces
of information are introduce into the repository.

Given the fact that the collaborative editing is not in question, the
solution for the problems of redundancy and incongruity must be
addressed by a powerful searching mechanism. Whenever a new piece of
information is about to be introduced, the repository must be searched
for an equal or similar piece of information, giving to the
collaborator the choice to give up (the ``new'' information is not new
after all), or update the ``old'' information with new facts.

\subsection{Technical Issues}
\label{sec:TechnicalIssues}

Connecting ``clients'' and ``servers'', i.e. geometric knowledge
manipulator applications (e.g DGS, GATP) and repositories of such
knowledge, can be done in two major ways.

\begin{itemize}
\item A one-to-one specific connection.

  For example we can use two systems, {\WGL} and {\TGTP}, and connect
  them in such a way that {\WGL} will send the query in the exact
  format that {\TGTP} is expecting and {\TGTP} will reply with an
  answer specifically tailored to {\WGL}.

  We can even think in a solution where the mechanisms (algorithms)
  implemented in one system are replicated in the other system. For
  example, opening the access to {\TGTP}'s database to the {\WGL}
  system and copying the search mechanisms of {\TGTP} to {\WGL}, with
  the smallest number of changes to make it work within {\WGL}.
  
\item A many-to-many generic connection.

  Like the \emph{Open Data Base Connectivity} standard (ODBC), the
  application programming interface (API) for accessing database
  management systems~\cite{Silberschatz2010}, a \emph{geometric API}
  is needed.

  The specification of such an API will allow to establish an easy (or
  at least standard) way to interconnect ``clients'' and ``servers''
  of geometric knowledge in a generic way. The existence of such an
  API would allow the interconnection between systems developed by
  different groups, systems with different aims, systems with
  different types of users, but with the common ``object'': geometry.
\end{itemize}

The one-to-one specific connection is just that. A connection specific
to two specific systems. For {\WGL} and {\TGTP} such task would be
easy, taking into account that the developer teams of both systems
overlap. For two systems developed by two different teams it will be,
in our opinion, a difficult negotiation to start with and, in the long
run, an almost impossible interconnection to maintain, whenever both
systems are still evolving.

In view of that, the many-to-many generic connection seems a more
sensible investment. What are the difficulties in establishing such an
API? What are the steps already done? What are the steps still to be
done?

When interconnecting two systems, a ``client'' and a ``server'' we
have to specify how the client will send the requests to the server
and how the server answers. Two distinct situations: the client wants
to query the server; the client wants to insert (or modify)
information in the server.

The first question has to do with searching and adaptive filtering.
The second with collaborative editing. In both cases the specification
of an API is needed.

The geometric API should specify:

\begin{itemize}
\item how to send text queries (simple or complex) to the server;
\item how to send geometric queries to the server;
\item how to send filters to the server;
\item how to combine filters and queries (filtered queries);
\item how to send users profiles (for the adaptive features in the
  server);
\item how to send new information, or update existing information, to
  the server;
\item the server's response.
\end{itemize}

In the following, the description of a first experiment in the
{\WGL}$\leftrightarrows${\TGTP} test-case will be describe.

\subsection{WGL Meets TGTP}
\label{sec:ClientQueryingServer}

The connection between clients and a server is a question of sending
the ``sentence'' or the geometric construction a user wants to query.

The text queries can be sent as a string. A format like \emph{JSON}, a
lightweight data-interchange
format,\footnote{\url{http://www.json.org/}} could be used to send the
query.

The geometric queries are more complex. As both {\WGL} and {\TGTP}
use the same DGS, \emph{GeoGebra}, the \texttt{ggb}
file could be sent by {\WGL}, received by {\TGTP}, processed and a
list of \texttt{ggb} files would be sent back to {\WGL}.  But, in a
more generic way, a common format like {\itogatp} should be used. The
{\itogatp} format~\cite{Quaresma2015a} is an extension of the {\itog}
(Intergeo) common
format\footnote{\url{http://i2geo.net}}~\cite{Santiago2010} that also
supports conjectures and proofs produced by geometric automatic theorem
provers. The {\itogatp} library is an open source
project,\footnote{\url{https://github.com/GeoTiles/libI2GATP}}
implemented in {\cpp}, to support the \itogatp\ common format. The
library implements methods to manage the {\itogatp} container and
filters between different GATP/DGS.



\paragraph{{\WGL} Querying {\TGTP}}
\label{sec:WGLqueryingTGTP}
In order to make the connection {\WGL}$\rightarrow${\TGTP}, allowing
{\WGL} access the {\TGTP} database, two applications were created, a
{\WGL} client and a {\TGTP} server.


The flow of information between {\WGL} and {\TGTP} is as follow.

\begin{enumerate}
\item In the \emph{List of Constructions} {\WGL}'s Web page, the
  {\WGL}'s user has a button \emph{Query the {\TGTP} server}, that
  opens a form, allowing the introduction of the query.
\item The {\WGL} client program is called (a \emph{system call}) with
  that query (string) as one of the command line
  arguments.\footnote{The command line call is ``\texttt{./clientWGL
      <TGTPserver\_name> <port> <query>}''}
\item The {\WGL} client will create a \emph{JSON} object with the
  query and the filter that asks for \emph{GeoGebra} constructions
  only. The \emph{JSON} object is then sent to the {\TGTP} server.

  The format of the \emph{JSON} object sent by the {\WGL} client is:
\begin{lstlisting}[frame=single,basicstyle=\small]
{
  "Query" : "<query>",
  "Filters" : "<filter> | <filter1 AND ... AND filterN>"
}                
\end{lstlisting}
  where the second component (``Filters'') is optional.
\item The {\TGTP} server receives the \emph{JSON} object, parses it,
  builds the \emph{SQL} query, sends it to the {\TGTP} database server,
  receives the answer, produces the corresponding \emph{JSON} object
  and sends it back to the {\WGL} client.

  The format of the \emph{JSON} object sent by the {\TGTP} server is:
\begin{lstlisting}[frame=single,basicstyle=\small]
{
  "Theorem Identifier1" : 
     "{"Name":"<name1>","Description":"<description1>","Code":"<code1>"
      ...,
  "Theorem IdentifierN" : 
      "{"Name":"<nameN>","Description":"<descriptionN>","Code":"<codeN>"
}                
\end{lstlisting}
  
\item The {\WGL} client receives the \emph{JSON} object (closing the
  network connection), parses it, builds a \emph{SQL} instruction
  corresponding to the insertion of the new construction(s) and sends it
  to the {\WGL} database server.
\item The {\WGL} Web page is then reloaded (automatically), updating
  the user's list of constructions.
\end{enumerate}

\begin{figure}[htb!]
  \centering
  \includegraphics[width=0.9\textwidth]{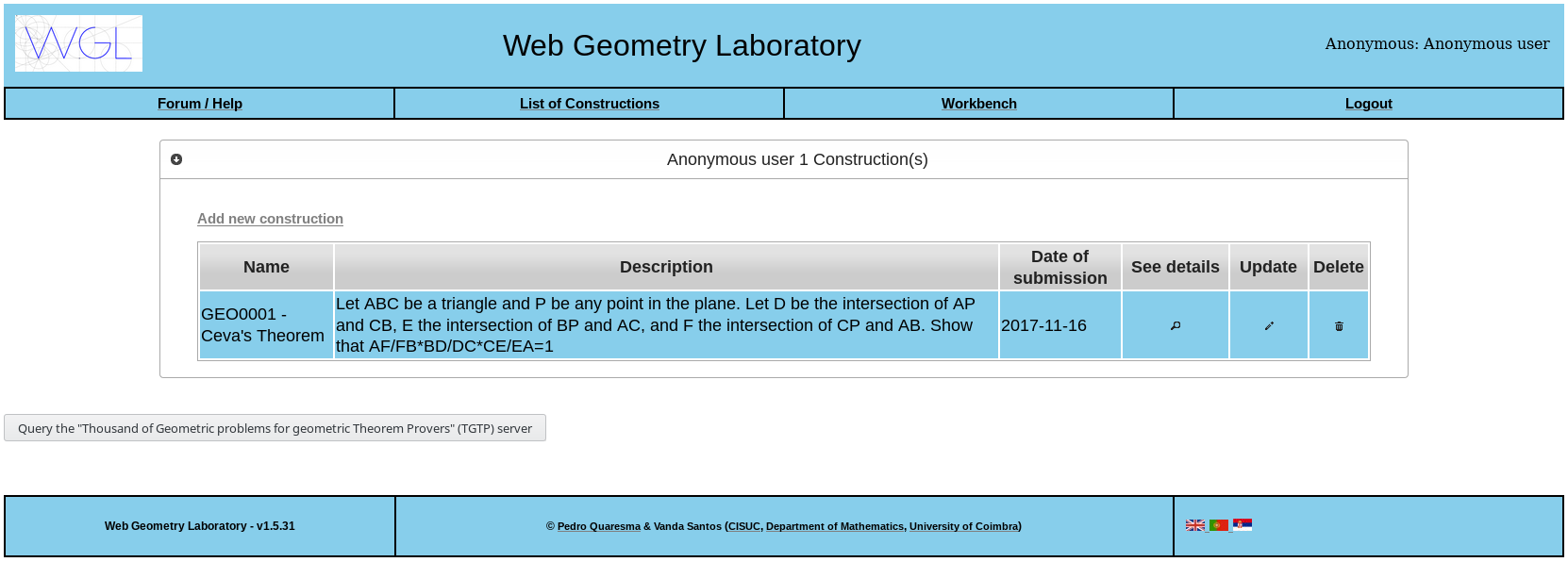}
  \caption{List of Constructions (after the query)}
  \label{fig:afterquery}
\end{figure}

Figure~\ref{fig:afterquery} show the anonymous user's list of
constructions, just after the query ``ceva'' was sent. The new
construction was incorporated in the previously empty list of
construction and it is now up to the user to make the best use of the
new information.

The server is an application that is listening, in an infinite cycle,
from a given socket (\emph{IP:port}). The clients are applications
that make requests to that socket, sending the query, and receiving
the answer. The current experimental implementation uses the anonymous
user of {\WGL} and the number of successful queries is limited because
we are looking for (native) \emph{GeoGebra} constructions in {\TGTP}
without using  (for now) the converters from other formats.



Both programs, \texttt{clientWGL} and \texttt{serverTGTP}, were written
in \emph{C++} with the libraries \emph{cppconn}, \emph{socket} and
\emph{ptree}, for the \emph{MySQL} connection, the network connection
between clients and servers and \emph{BOOST} property trees library
and \emph{JSON} parser methods, respectively.

The client program is an open source program, contained in the {\TGTP}
project.\footnote{\url{https://github.com/GeoTiles/TGTP}} Anyone
can write clients to access the {\TGTP} server.

\paragraph{{\TGTP} Enlarged Users Base}
\label{sec:tgtpenlargeuserbase}

Assuming that the repository (server) uses the {\itogatp} format, any
insertion and/or updating should be done by building the container, on
the client side, and afterwards send it to the server. As the {\itogatp}
associated library has methods to manipulate the {\itogatp}
containers, it should be a matter of collecting all the information
needed and then, using the appropriated methods, build the container
and send it to the server.

Using, again, the test-case as an example: {\TGTP} already uses the
{\itogatp} format to store the problems in its database; from the
{\WGL} side it would be a question of collecting all the information
needed, build the {\itogatp} container using the methods in the
{\itogatp} library and send the information to the {\TGTP}
server. This is still to be implemented.

\section{Conclusions and Future Work}
\label{sec:conclusions}

Instead of a giant (heavy and difficult to use and maintain) tool,
trying to cover all features of the many specialised tools, the
interconnection of those specialised tools seems much more promising.
The connection between {\WGL} and {\TGTP} is only an example of the
connections that should be made between geometric tools. Efforts like
\emph{OpenGeoProver}, a open-source project of a library of
GATPs,\footnote{\url{https://github.com/ivan-z-petrovic/open-geo-prover/}}
to be used by different systems
(e.g. \emph{GeoGebra}~\cite{Botana2015a}), or the connection between
DGS and CAS~\cite{Escribano2010}, share the common approach of
``mosaic tool'', a tool composed of many small pieces forming a
beautiful picture.

The main challenge faced by developers, when trying to connect
different types of systems, is the definition (and implementation) of
common formats. Having the {\itogatp} common format as a starting
point and the corresponding open source library \texttt{libI2GATP} as
a form to manipulate the {\itogatp} container, we should enlarge both
as needed to allow different systems to exchange information.

The definition of an \emph{ODBC} counterpart for geometry would allow
to have a layer between clients and servers, isolating the nasty
details from the developer and thus allowing an easy interconnection
between systems.

The other big challenge is the retrieving of information, i.e. queries
sent from the clients to the servers. The definition of a query
language (an \emph{SQL} counterpart) for geometry, allowing to explore
text and geometric queries is needed.

Another issue is implementing adaptive features. Again a specification
of what are the students of geometry learning profiles should be
pursued.


We feel that having identified the main issues and having already done
some steps, it is, now, a question of filling in the gaps to make it
possible.

\bibliographystyle{eptcs}
\bibliography{eptcs2018PQVSNB}

\end{document}